\documentclass[aps,prl,reprint,superscriptaddress,showpacs]{revtex4-1}

\usepackage[]{hyperref} 
\usepackage{amsmath}
\usepackage{amsfonts}
\usepackage{textcomp}
\usepackage[]{graphicx}
\usepackage{epstopdf}
\usepackage{mathrsfs}
\usepackage{amsbsy}
\usepackage{cleveref}
\usepackage[rightcaption]{sidecap}

\crefformat{equation}{Eq.~(#2#1#3)}
\crefformat{figure}{Fig.~#2#1#3}

\Crefformat{equation}{Equation~(#2#1#3)}
\Crefformat{figure}{Figure~#2#1#3}

\hypersetup{pdfauthor={Ma, Walasik, Litchinitser},pdftitle={Linear PT PRL}}

\begin{document}

\title{Meta-$\mathcal{PT}$ Symmetry in Asymmetric Directional Couplers}

\author{Chicheng Ma}
\affiliation{Department of Electrical Engineering, University at Buffalo, The State University of New York, Buffalo, New York 14260, USA}
\author{Wiktor Walasik}
\affiliation{Department of Electrical Engineering, University at Buffalo, The State University of New York, Buffalo, New York 14260, USA}
\author{Natalia M. Litchinitser}
\affiliation{Department of Electrical Engineering, University at Buffalo, The State University of New York, Buffalo, New York 14260, USA}
\email[]{wiktorwa@buffalo.edu}

\date{\today}

\begin{abstract}
Despite the benefits that directional coupler based parity-time symmetric systems may offer to the field of integrated optics, the realization of such couplers relies on rather strict design constraints on the waveguide parameters. Here, we investigate directional couplers built of asymmetric waveguides that do not fulfill parity-time symmetry. Nevertheless, we demonstrate that for properly designed parameters, at least one mode of such couplers shows a behavior similar to the one observed in parity-symmetric systems. We find an analytical condition relating gain and loss that enables such a behavior. Moreover, if the individual waveguides composing the asymmetric coupler are designed such that the propagation constants of their modes are identical, the behavior of both super-modes supported by the coupler resembles that of the parity-time symmetric systems.
\end{abstract}


\maketitle

Directional optical couplers have attracted a lot of attention in the fields of integrated and fiber optics owing to their versatile applications, including spectral filters, optical switches, signal multiplexers and modulators, power dividers, and optical cross-connects. Performance of directional couplers as a function of waveguide size, spacing, refractive index profiles, and length has been studied in detail in Ref.~\cite{dietrich1991theory}. Moreover, directional optical couplers with gain, loss, and nonlinear effects have been analyzed and demonstrated experimentally~\cite{finlayson1990spatial,setterlind1986directional,litchinitser2007optical,musslimani2008optical}.

Recently, studies of so-called parity-time symmetric~\cite{Bender98,Bender02,Bender07} ($\mathcal{PT}$-symmetric) directional couplers attracted a lot of attention~\cite{ruschhaupt2005physical,El-Ganainy07,makris2008beam,klaiman2008visualization,Bendix09,guo2009observation,ruter2010observation,peng2014parity,choi2015parity}. $\mathcal{PT}$-symmetric systems belong to a more general class of pseudo-Hermitian systems~\cite{mostafazadeh2002pseudo,mostafazadeh2002pseudo2}, which despite of being non-Hermitian may still possess real eigenvalues. In optics, $\mathcal{PT}$ symmetry relies on a proper modulation of gain and loss, and was predicted to enable unconventional switching and memory functionalities. In particular, it was shown that a $\mathcal{PT}$-symmetric phase transition occurs from a full $\mathcal{PT}$-symmetric regime (both modes with real propagation constants) to a broken $\mathcal{PT}$-symmetric regime, where one mode is amplified while the other is attenuated~\cite{El-Ganainy07,klaiman2008visualization}. Change from full to broken $\mathcal{PT}$ regime occurs with the increase of gain and loss in the system at a specific transition point.
Both linear~\cite{ruschhaupt2005physical,El-Ganainy07,makris2008beam,klaiman2008visualization,Bendix09,guo2009observation,ruter2010observation,peng2014parity,choi2015parity} and nonlinear~\cite{Driben11,barashenkov2012breathers,Driben12,lumer2013nonlinearly,bludov2014pt,Walasik15} $\mathcal{PT}$-symmetric waveguide systems were analyzed and many remarkable phenomena were discovered, such as suppression of time reversal~\cite{sukhorukov2010nonlinear} and unidirectional nonlinear $\mathcal{PT}$-symmetric coupling~\cite{ramezani2010unidirectional}.   

Until now, the majority of studies focused on systems strictly fulfilling the $\mathcal{PT}$-symmetry condition: $\epsilon(x) = \epsilon^*(-x)$, implying waveguides with the same size, symmetric distribution of the real part of permittivity, and anti-symmetric gain/loss profile~\cite{ruter2010observation,El-Ganainy07}. These are rather difficult constraints to satisfy in laboratory experiments. Therefore, in this paper, we study asymmetric couplers with dissimilar cores and find a $\mathcal{PT}$-like regime of wave propagation in such systems. In particular, we show that even an asymmetric coupler with complex permittivity profile, provided gain and loss are properly balanced, supports a mode with a real propagation constant. In such a coupler, the transition point that separates the regime in which a particular mode nearly conserves the energy from that, in which it experiences gain or loss is analytically identified. For values of gain and loss below this point, the mode experiences a very low attenuation, negligible at the sub-millimeter propagation distances typical for integrated optical systems. Above this point, the mode experiences gain that grows rapidly with the increase of the magnitude of the imaginary part of permittivity. Moreover, we show that in a special case, when the propagation constant of the isolated waveguides building the asymmetric coupler are equal, the behavior of both super-modes supported by the coupler resembles that of the $\mathcal{PT}$-symmetric systems. 

\begin{figure}[!b]
	\includegraphics[width = \columnwidth, clip=true, trim = {0 0 0 0}]{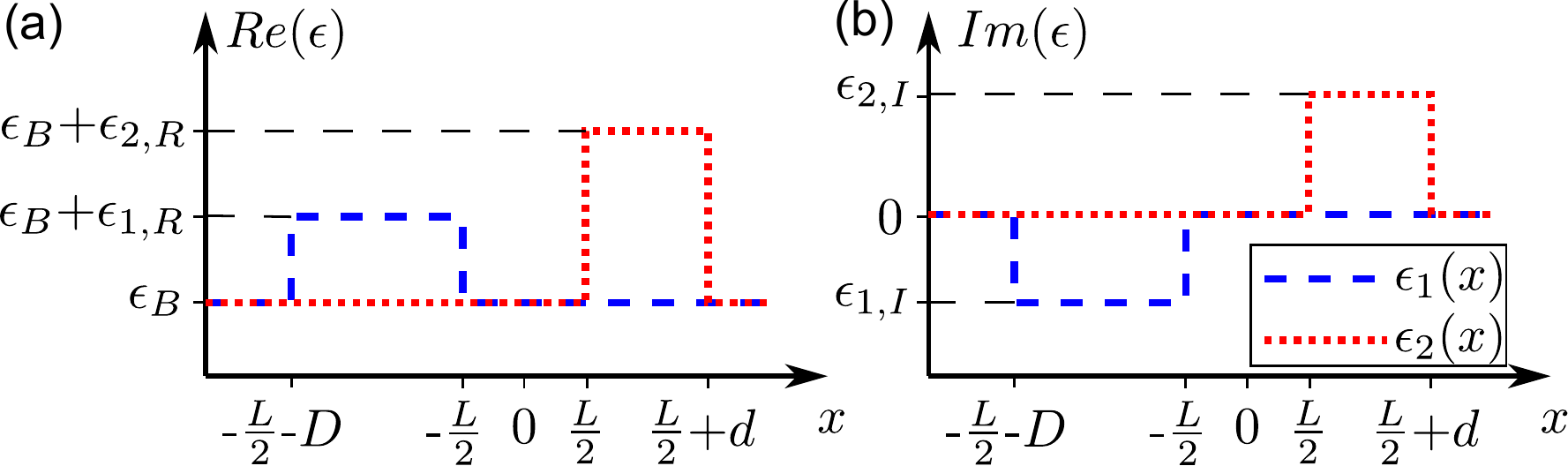}
	\caption{Geometry of the studied asymmetric coupler. (a) Real and (b) imaginary part of permittivity profile of the first (blue dashed) and the second (red dotted) waveguide.}
	\label{fig:geom}
\end{figure}

The permittivity profile of the one-dimensional (1D) linear directional coupler studied here is given by
\begin{equation}
\epsilon(x) = \epsilon_1(x) + \epsilon_2(x) - \epsilon_B,
\label{eqn:perm_coupler}
\end{equation}
and is shown in \cref{fig:geom}. Here, $\epsilon_B$ denotes the value of the relative background permittivity and $\epsilon_1(x)$ and $\epsilon_2(x)$ are the complex permittivity profiles of the first and the second waveguide, respectively, The opto-geometric parameters of the two waveguides are indicated in \cref{fig:geom}. One of the waveguides introduces loss to the system ($\epsilon_{1,I}<0$), while the other one provides gain ($\epsilon_{2,I}>0$).

In order to study light propagation in the system described by \cref{eqn:perm_coupler}, we solve the scalar wave equation for the electric field $E$:
\begin{equation}
\nabla^2 E(x,z) + k_0^2 {\epsilon}(x)  E(x,z) = 0,
\label{eqn:wave}
\end{equation}
where $k_0 = 2\pi/\lambda$ denotes the free-space wavevector and $\lambda$ is the free-space wavelength of light. The operator $\nabla^2 = \partial^2/\partial x^2 + \partial^2/\partial z^2$ denotes the 2D Laplacian and the wave propagates in the $z$-direction. \Cref{eqn:wave} is obtained from Maxwell's equations under the assumption that  both the structure and the field distributions are invariant along the $y$-direction.

First we analyze \cref{eqn:wave} using the coupled mode theory~\cite{sukhorukov2010nonlinear}. In this case, the electric field in the coupler can be written as a sum of the fields of the individual waveguides $E(x,z) = \sum_{i=1}^{2} \tilde{A}_i(z) \psi_i(x)$, where $\psi_i(x)$ denotes the transverse field distribution of the (fundamental) mode in the isolated waveguide described by the permittivity distribution $\Re e[\epsilon_{i}(x)]$. The fast varying envelope of the corresponding mode is denoted by $\tilde{A}_i(z)$.

Using the slowly varying envelope approximation (SVEA) $\tilde{A}_i(z) = A_i(z) e^{-j\beta_i z}$ and neglecting small overlap terms, \cref{eqn:wave} can be transformed into coupled mode equations (CME) for a directional coupler with gain and loss:
\begin{equation}
j\frac{\mathrm{d} \tilde{A}_i}{\mathrm{d}z} = (\beta_i + C_i +j \gamma_i)\tilde{A}_i + \kappa_i \tilde{A}_{3-i},
\label{eqn:coupled}
\end{equation}
where $i \in \{1,2\}$, and $\gamma_1$ ($\gamma_2$) is loss (gain) coefficient. $C_i$ and $\kappa_i$ denote self-coupling and  coupling coefficients, respectively. These coefficients are calculated as:
\begin{subequations}
	\begin{align}
	C_i &= N_i \epsilon_{3-i,R}  \langle \psi_i| \psi_i \rangle_{WG(3-i)},\\
	\gamma_i &= N_i  \epsilon_{i,I} \langle \psi_i| \psi_i \rangle_{WGi}, \label{eqn:gamma} \\
	\kappa_i &= N_i \epsilon_{i,R} \langle \psi_{3-i}| \psi_i \rangle_{WGi}.
	\end{align}
\end{subequations}
Here $N_i = k_0^2/[2\beta_i \int_{-\infty}^{\infty} \psi_i(x)\psi_i^*(x)\mathrm{d}x]$, and $\langle f| g \rangle_{WGi} = \int_{WGi} f(x) g^*(x) \mathrm{d}x$. The integral limits $WGi$ denote integration over the cross-section of the $i$th waveguide.

Dynamical properties of the asymmetric coupler can be found analytically from \cref{eqn:coupled}. To this end, we assume that the propagation constants of the two modes are nearly the same $\beta_1 \approx \beta_2 \approx \beta$, such that the solution can be written as $\tilde{\mathbf{A}} = \mathbf{A}e^{-j\beta z}$, where $\tilde{\mathbf{A}} = [\tilde{A}_1, \tilde{A}_2]^T$ and $\mathbf{A} = [A_1, A_2]^T$. This allows us to rewrite \cref{eqn:coupled} in a matrix form $H \mathbf{A} = \beta \mathbf{A}$, where the Hamiltonian is given by
\begin{equation}
H =
\begin{pmatrix}
h_{11} & h_{12} \\ 
h_{21} & h_{22}
\end{pmatrix} 
=
\begin{pmatrix}
\beta_1 + C_1 + j\gamma_1 & \kappa_1 \\ 
\kappa_2 & \beta_2 + C_2 + j\gamma_2 
\end{pmatrix}.
\end{equation}
The eigenvalues $\beta$ correspond to the propagations constant of the coupler super-modes. Nontrivial solutions of \cref{eqn:coupled} exist only if $\det(H-\beta\mathbf{I}) = 0$. This condition allows us to find the propagation constants of the coupler modes to be:
\begin{align}
\beta_{\pm} =(h_{11} + h_{22} \pm \sqrt{ (h_{11} + h_{22})^2 - 4 \det(H) })/2.
 \label{eqn:beta} 
\end{align}

The dynamic properties of the studied coupler are determined by \cref{eqn:beta} and depend on the coupling strength, self-coupling coefficients, and gain and loss coefficients~\cite{syms1992optical}. In the general case, the propagation constants given by \cref{eqn:beta} are complex. However, if the parameters of the waveguides are carefully chosen, the imaginary part of one of the propagation constants can be eliminated. Indeed, the condition for a purely real propagation constant is obtained from \cref{eqn:beta}:
\begin{equation}
\Im m(\beta_{\pm}) = [(\gamma_1+\gamma_2)\pm q]/2 = 0,
\label{eqn:condition}
\end{equation}
where $q = \mathrm{sgn}(b)/ [(\sqrt{a^2 + b^2} -a)/2]^{1/2}$, and
\begin{subequations}
	\begin{align}
	a &= \{(\beta_1+ \beta_2 + C_1 + C_2)^2 - (\gamma_1 + \gamma_2)^2\} \nonumber \\ &+ 4 \kappa_1\kappa_2 - 4[(\beta_1+C_1)(\beta_2+C_2) -\gamma_1\gamma_2],\\
	b &= 2 (\beta_1+\beta_2 + C_1 +C_2)(\gamma_1 + \gamma_2)\nonumber \\ &- 4[(\beta_1+C_1)\gamma_2 + (\beta_2+C_2)\gamma_1].
	\end{align}
\end{subequations}
The choice of the $+$ ($-$) sign in \cref{eqn:condition} corresponds to the lossless fundamental (second-order) mode. The system can be optimized for only one mode that conserves its energy during the propagation at a time. Here, we optimize the system in such a way that the fundamental modes is lossless, and therefore we choose the $+$ sign in all further computations.

For fixed geometry and real permittivity profile of the coupler, Eqs.~(\ref{eqn:condition}) and (\ref{eqn:gamma}) yield a relation between gain $\epsilon_{2,I}$ and loss $\epsilon_{1,I}$ for which we obtain a mode with a real propagation constant in a not $\mathcal{PT}$-symmetric coupler. Such systems will be called thereafter meta-$\mathcal{PT}$ (m-$\mathcal{PT}$) systems. Below we present a design of an asymmetric m-$\mathcal{PT}$ coupler and compare it to a $\mathcal{PT}$-symmetric coupler with similar parameters.  

The parameters of our asymmetric m-$\mathcal{PT}$ coupler are the following (see \cref{fig:geom} for definitions): $\epsilon_B = 3$, $\epsilon_{1,R} = 0.25$, $\epsilon_{1,I} = -0.02$, $D=0.7$~$\mu$m, $\epsilon_{2,R} = 0.35$, $d = 0.5$~$\mu$m and the separation between waveguides is $L=0.2$~$\mu$m. At the wavelength of light used ($\lambda = 0.8$~$\mu$m) both waveguides support only one mode. To complete the design, the value of gain has to be chosen according to \cref{eqn:condition}. This results in  $\epsilon_{2,I} = -M\epsilon_{1,I}$, where $M$ denotes an adjustment factor we introduce in order to balance gain and loss in the system and achieve a real propagation constant of the fundamental mode. This analytical CME-based approach yields $M = 0.4729$.  

\begin{figure}[!t]
	\includegraphics[width = 0.49\columnwidth, clip=true, trim = {50 225 95 245}]{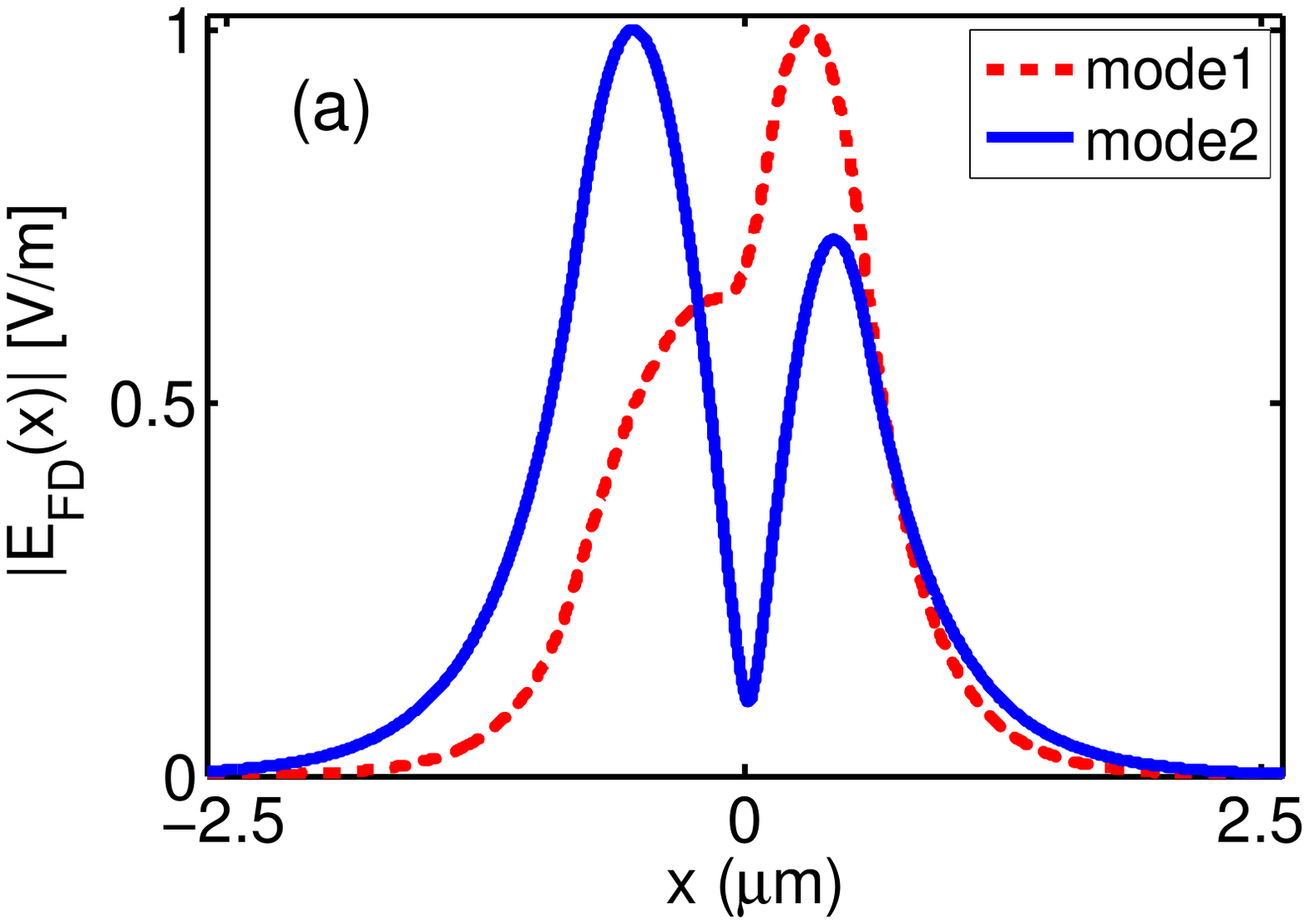}
	\includegraphics[width = 0.49\columnwidth, clip=true, trim = {50 225 95 245}]{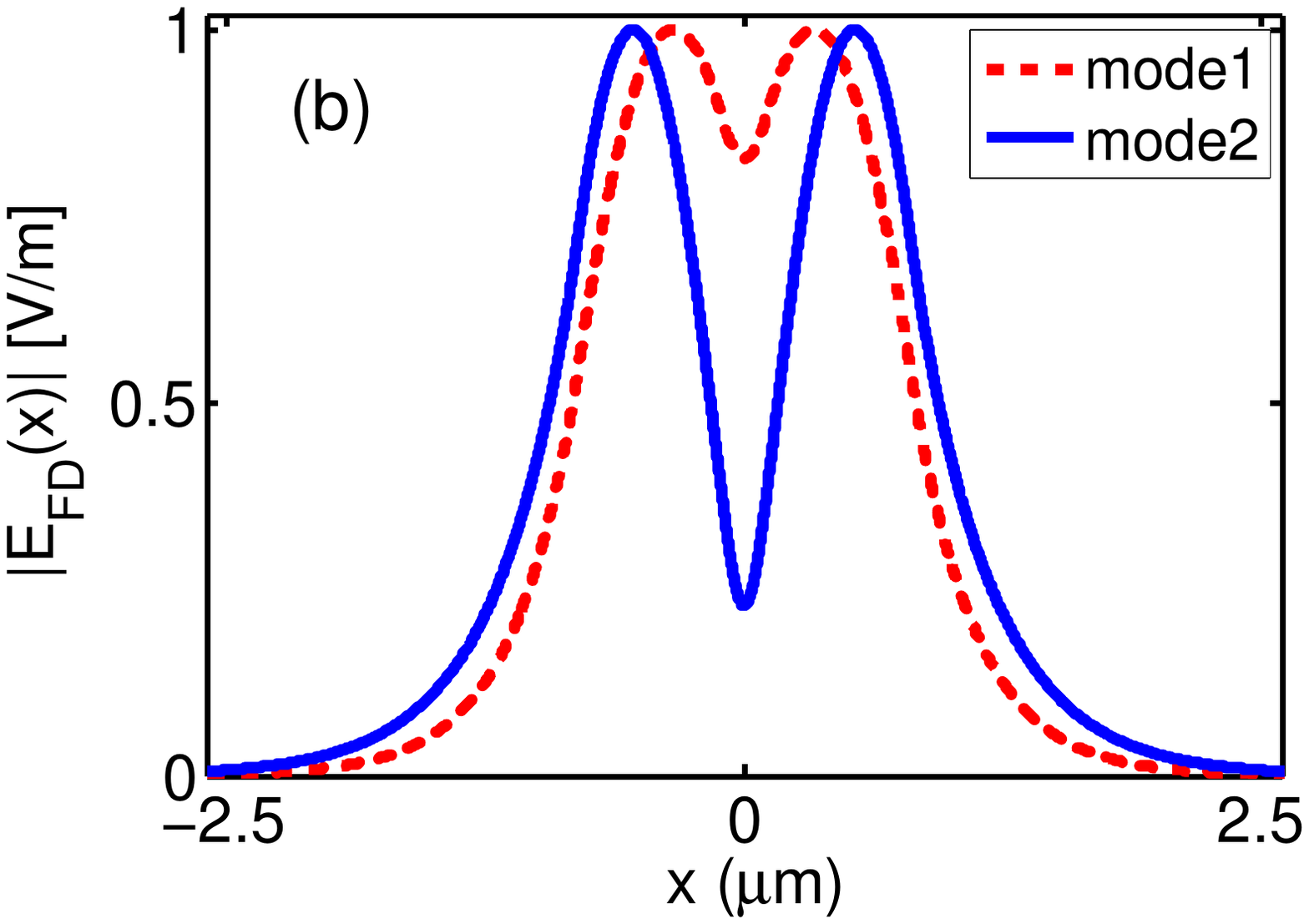}\\
	\includegraphics[width = 0.49\columnwidth, clip=true, trim = {-16 0 0 0}]{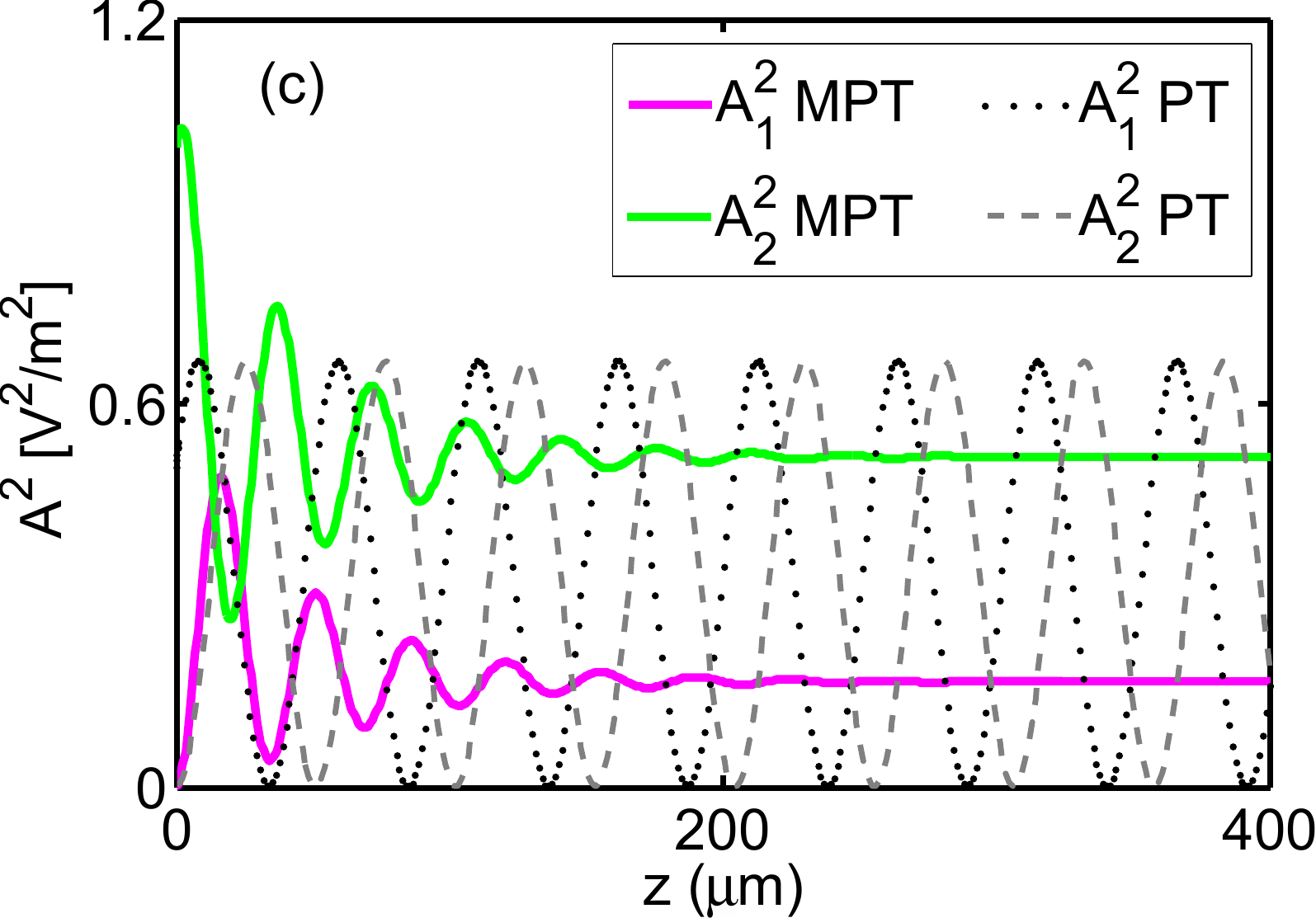}
	\includegraphics[width = 0.49\columnwidth, clip=true, trim = {0 0 40 0}]{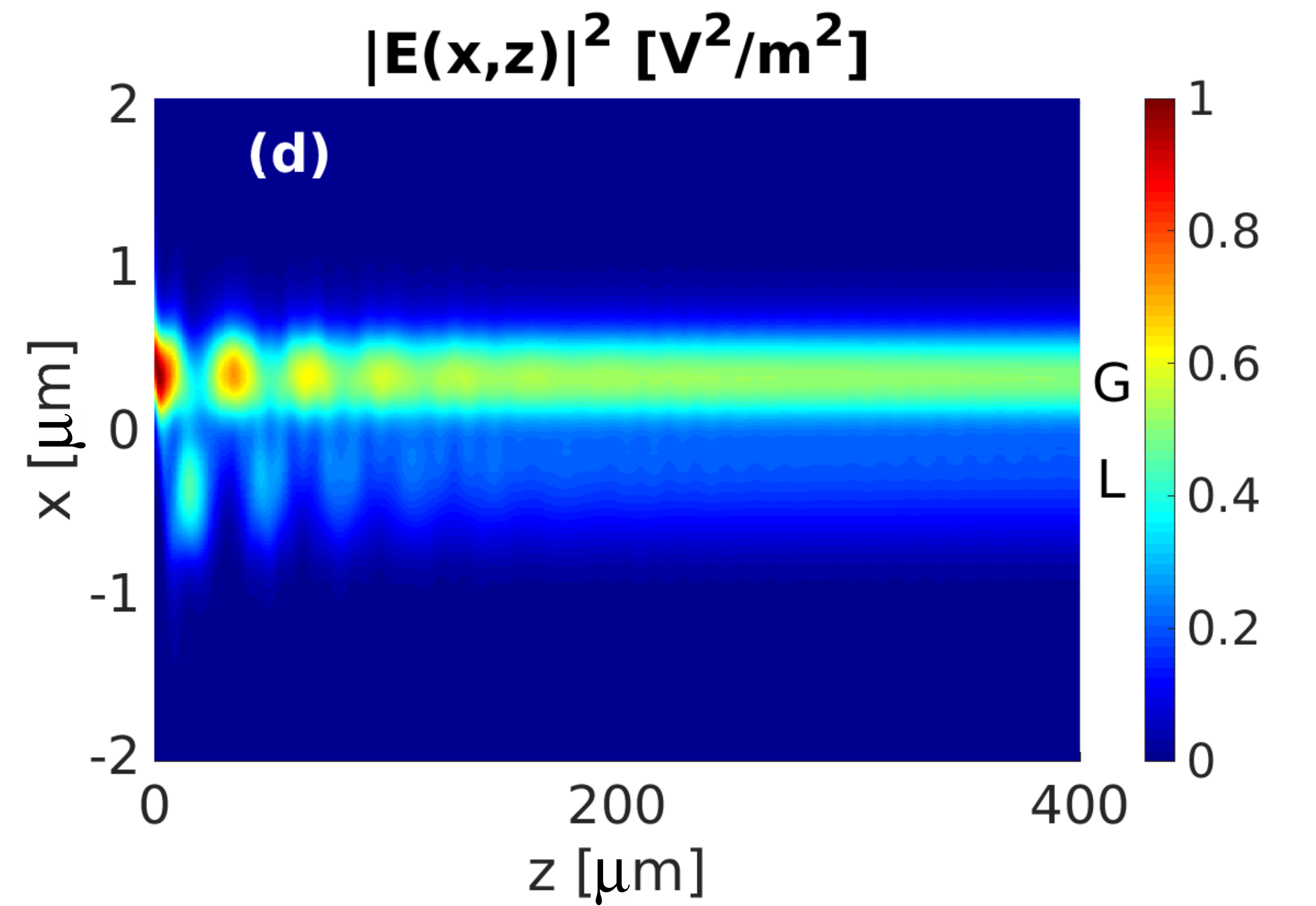}\\
	\includegraphics[width = 0.49\columnwidth, clip=true, trim = {50 225 85 235}]{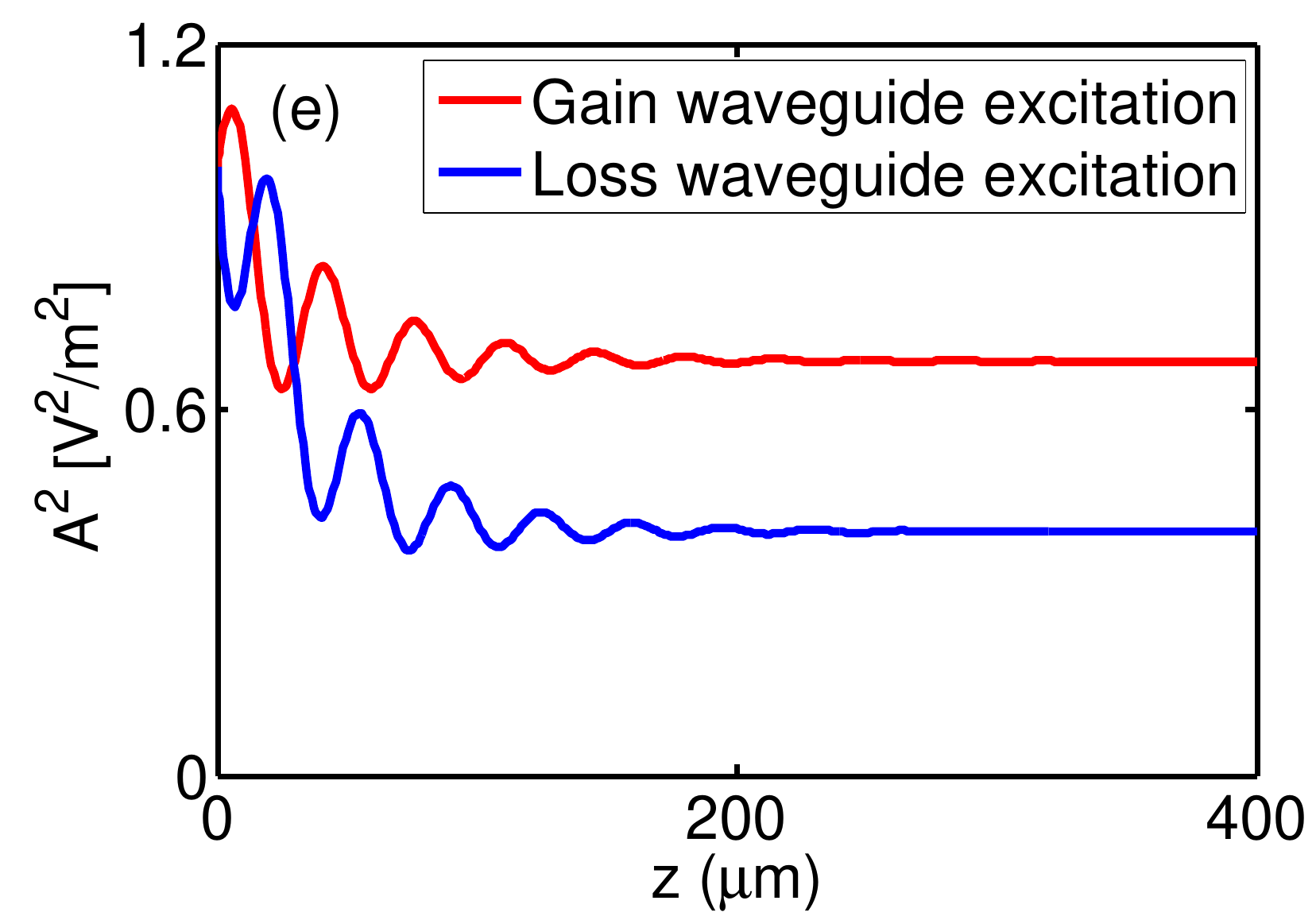}
	\includegraphics[width = 0.49\columnwidth, clip=true, trim = {0 0 0 0}]{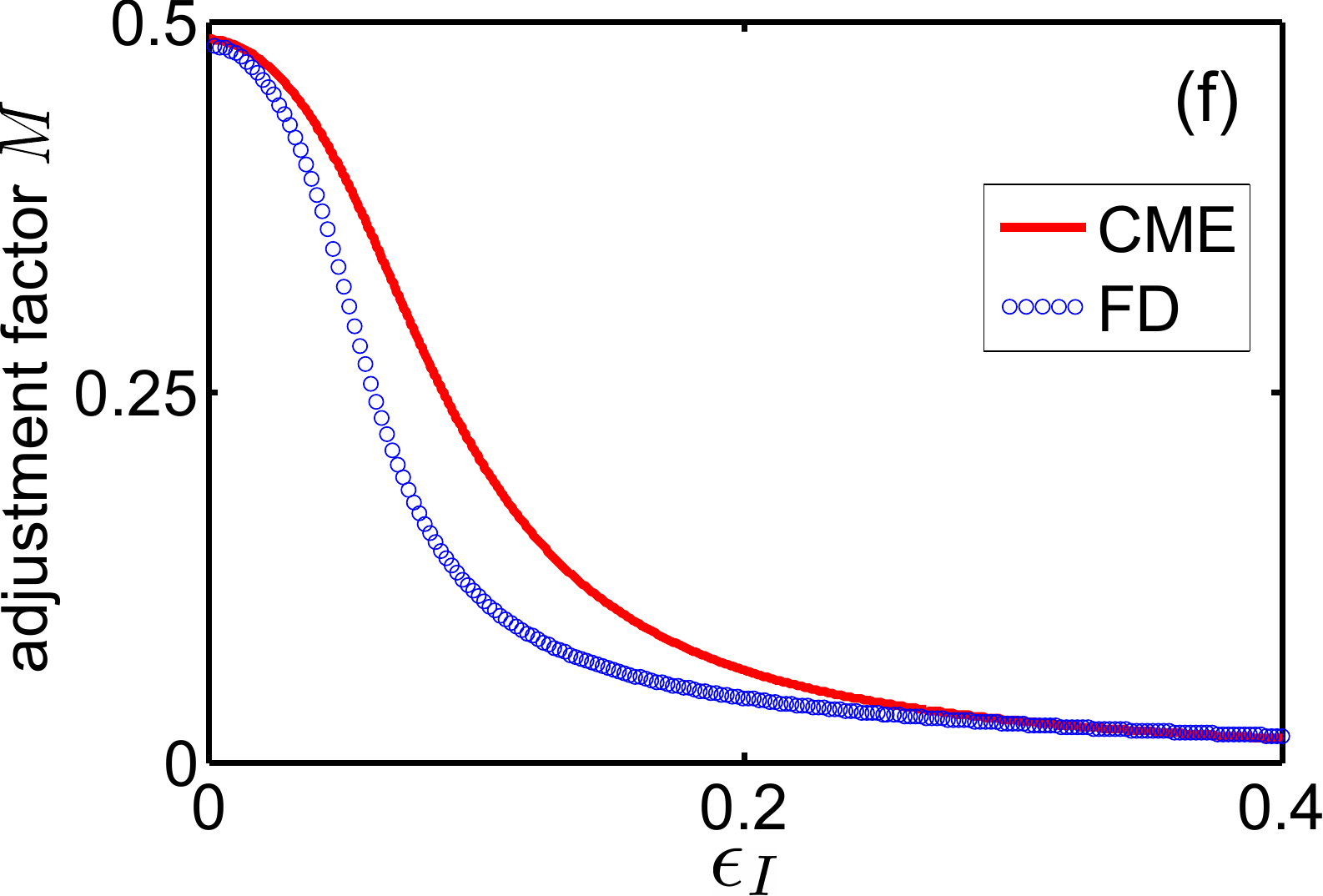}\\
	\caption{(a), (b) Mode profiles in (a) m-$\mathcal{PT}$ and (b) $\mathcal{PT}$-symmetric system calculated by using the FD method (red dashed curve---fundamental mode and blue solid curve---second-order mode). (c) Evolution of the slowly varying amplitudes $A_1^2$ (loss waveguide) and $A_2^2$ (gain waveguide) in m-$\mathcal{PT}$ coupler (solid color curves) and $\mathcal{PT}$-symmetric coupler (gray curves). (d) Evolution of the electric field $|E(x,z)|^2$ obtained from FFT-BPM (G and L denote gain ans loss waveguide, respectively). (e) Sum of the slowly varying amplitudes $A_1^2+A_2^2$ in  m-$\mathcal{PT}$ coupler for gain or loss waveguide excitation. (f) Adjustment factor $M$ as a function of $\epsilon_I$ under the condition of $\Im m(\beta_+) = 0$.}
	\label{fig:propagation}
\end{figure}

In order to confirm the validity of the analytical approach, we study the properties of our asymmetric coupler using a numerical method in which the modes are found by solving a $z$-independent Helmholz equation [obtained from \cref{eqn:wave}] using the Finite Difference (FD) method. This approach allows for an alternative and more physically intuitive way of finding the adjustment factor. We require the product of the absolute value of the imaginary part of the permittivity and the power propagating in each of the waveguides to be the same, which results in $M_{\textrm{FD}} = |\epsilon_{2,I}|/|\epsilon_{1,I}| = [\int_{WG1} |E_{\textrm{FD}}(x)|^2 \mathrm{d}x]/[\int_{WG2} |E_{\textrm{FD}}(x)|^2 \mathrm{d}x]$, where $E_{\textrm{FD}}(x)$ denotes the electric field distribution of the coupler mode found using the FD method presented in \cref{fig:propagation}(a). Calculations using this approach result in $M_{\textrm{FD}} = 0.4703$ for the fundamental mode, which is in a good agreement with the result obtained using the analytical method. In a conventional $\mathcal{PT}$-symmetric coupler, the adjustment factor $M = 1$, due to symmetry of the system [see mode profiles in \cref{fig:propagation}(b)]. In the following, all the results are obtained using the CME method and FD method. They are generally in a very good agreement; and therefore, we will only comment on the origin of any differences observed in the results.

For the optimized parameters of the coupler, it supports two modes shown in \cref{fig:propagation}(a): (i) the fundamental mode with the propagation constant $\beta_+ = 14.08$~$\mu$m$^{-1}$ and (ii) the second-order mode with  $\beta_- = (13.90 - j0.02)$~$\mu$m$^{-1}$. As intended by the design, the fundamental mode has a real propagation constant, while the second-order mode experiences loss.

Let us now study the light propagation in the m-$\mathcal{PT}$ coupler. The evolution of the slowly varying envelopes~$A^2_i$, computed by solving \cref{eqn:coupled} using the fourth-order Runge-Kutta algorithm~\cite{dormand1980family}, is shown in \cref{fig:propagation}(c). Here, we launch the light in the gain waveguide only. This excitation light overlaps with both the fundamental and the second-order coupler mode [shown in \cref{fig:propagation}(a)] and excites them both. These modes propagate in the coupler and their interference results in oscillations of light intensity in the waveguides. The period of the interference pattern can be estimated as $2\pi/\Re e(\beta_+ - \beta_-) \approx 35$~$\mu$m. Similar oscillations are present in standard $\mathcal{PT}$-symmetric systems (here with parameters $\epsilon_{1,R} = \epsilon_{2,R} =0.25$, $\epsilon_{1,I} = -\epsilon_{2,I} =  -0.02$, and $D=d=0.7$~$\mu$m,) upon a single waveguide excitation [see \cref{fig:propagation}(b) for mode profiles and gray curves in \cref{fig:propagation}(c) for slowly varying amplitude evolution]. In the full $\mathcal{PT}$-symmetric regime, these oscillations continue throughout the entire propagation as both modes have real propagation constants. On the contrary, in the m-$\mathcal{PT}$ systems, one mode experiences loss and decays at a characteristic length of $L_D = 1/\Im m(\beta_-) \approx 50$~$\mu$m. After several $L_D$'s, this lossy mode vanishes and for $z>200$~$\mu$m, we observe a stable propagation of the lossless fundamental mode, without the oscillations induced by interference. The dynamical behavior of the system obtained from CMEs is confirmed by numerical solution of a (1+1)D Schr\"odinger-like equation [obtained using SVEA in \cref{eqn:wave}] using Fast-Fourier Transform Beam-Propagation Method (FFT-BPM)~\cite{feit1978light} [see \cref{fig:propagation}(d)].

A very similar behavior is obtained if the loss waveguide is excited instead of the gain one. The only difference is the lower total intensity reached after the stabilization phase, as shown in \cref{fig:propagation}(e). After the stabilization the field remains confined in the fundamental mode, which means that the intensity ratio between the two waveguides is the same, regardless of which waveguide is excited. This result may offer a new functionality for all-optical switches. 

Let us now investigate the ratio between gain and loss $M=|\epsilon_{2,I}|/|\epsilon_{1,I}|$ required for the propagation constant of the fundamental mode to be purely real in the m-$\mathcal{PT}$ system with a fixed optogeometric parameters, but with various amplitudes of the imaginary part of the permittivity $\epsilon_I$ ($\epsilon_{1,I} = -\epsilon_I$, $\epsilon_{2,I} = M\epsilon_I$). \Cref{fig:propagation}(f) shows the dependence of the adjustment factor $M$ on $\epsilon_I$ calculated by CME method and the FD method. The maximum value of adjustment factor $M=0.49$ is obtained when the imaginary part of permittivity approaches zero. With the increase of $\epsilon_I$ the adjustment factor $M$ decreases rapidly. This means that the field distribution of the fundamental coupler mode becomes more asymmetric (more localized in the gain waveguide) and lower gain is needed in order to support stable propagation of this mode. The differences between the results of the CME and the FD methods are caused by the fact that the FD method, as opposed to the CME method, takes into account the imaginary part of the permittivity in calculations of the mode profiles.

$\mathcal{PT}$-symmetric systems exist in two regimes depending on the amplitude of gain and loss  $\epsilon_I$: full $\mathcal{PT}$-symmetric regime for low values of $\epsilon_I$ and broken $\mathcal{PT}$-symmetric regime for $\epsilon_I$ above the threshold value corresponding to the exceptional point. With the increase of $\epsilon_I$, the two real propagation constants of lossless modes of the full $\mathcal{PT}$-symmetric system become closer and coalesce at the exceptional point (see for example, Fig.~2 in Ref.\cite{klaiman2008visualization}). Above this point, two modes with complex conjugate propagation constants exist. 

\begin{figure}[!t]
	\includegraphics[width = 0.49\columnwidth, clip=true, trim = {0	0 0 0}]{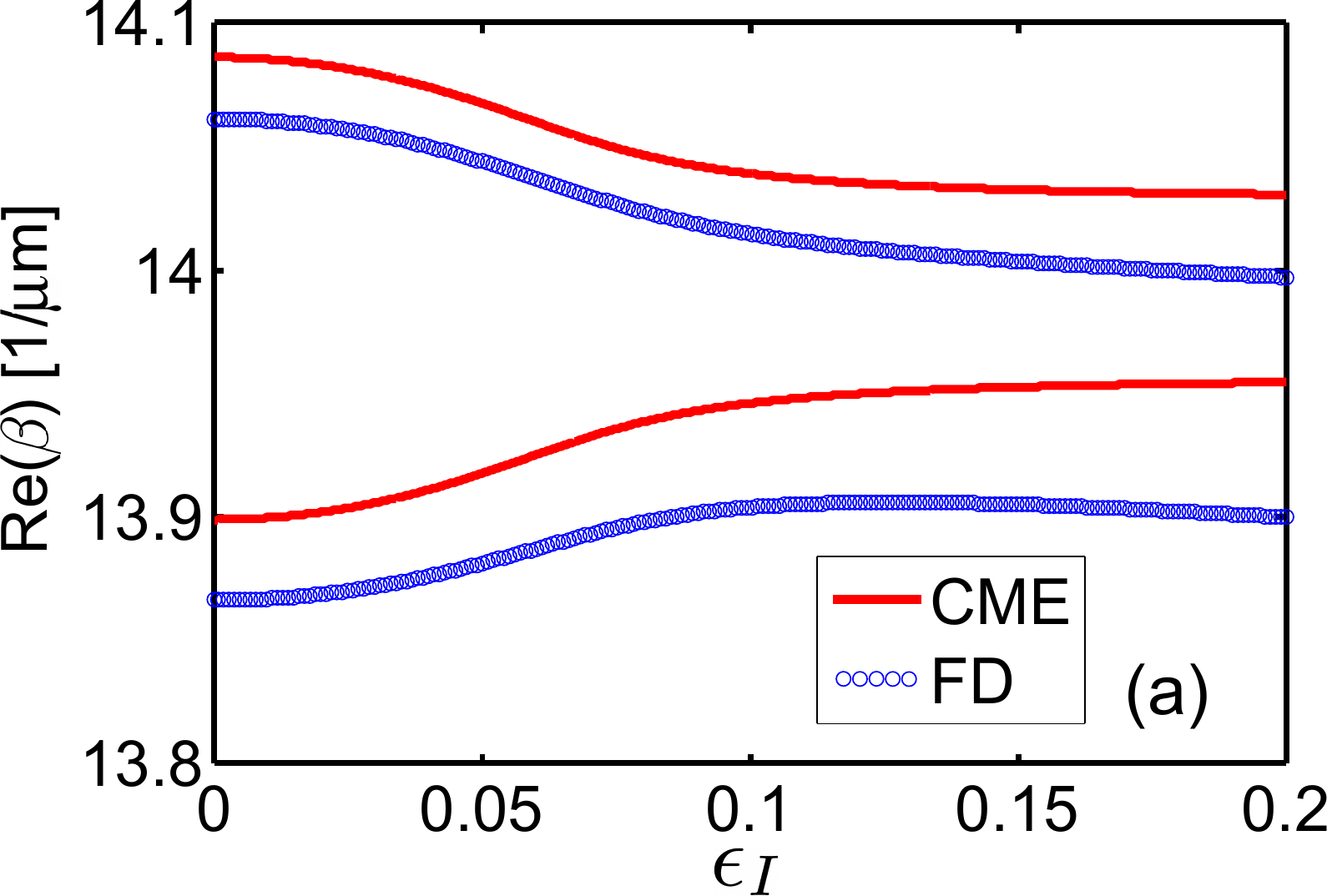}
	\includegraphics[width = 0.49\columnwidth, clip=true, trim = {0	0 0 0}]{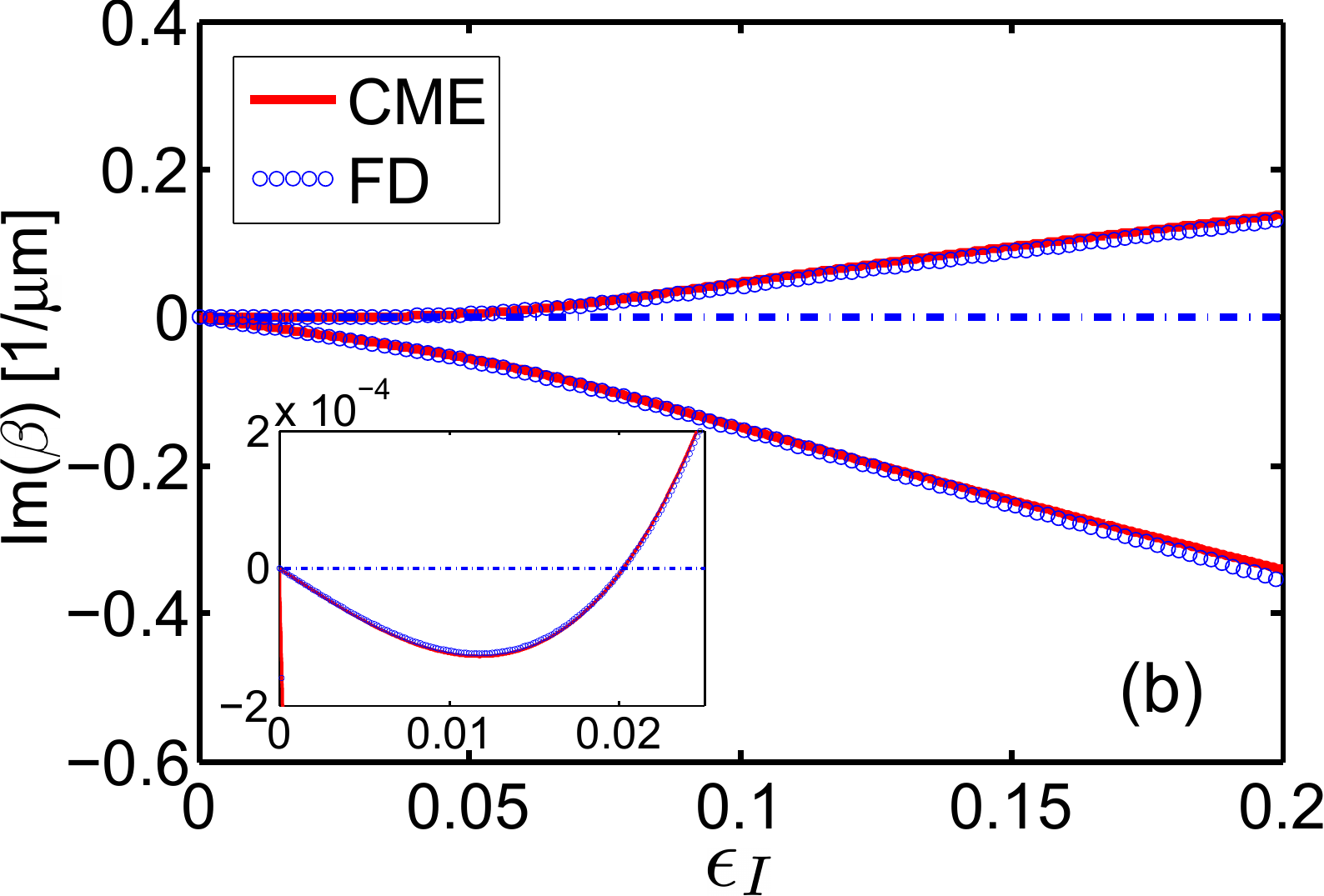}\\
	\includegraphics[width = 0.49\columnwidth, clip=true, trim = {0	0 0 0}]{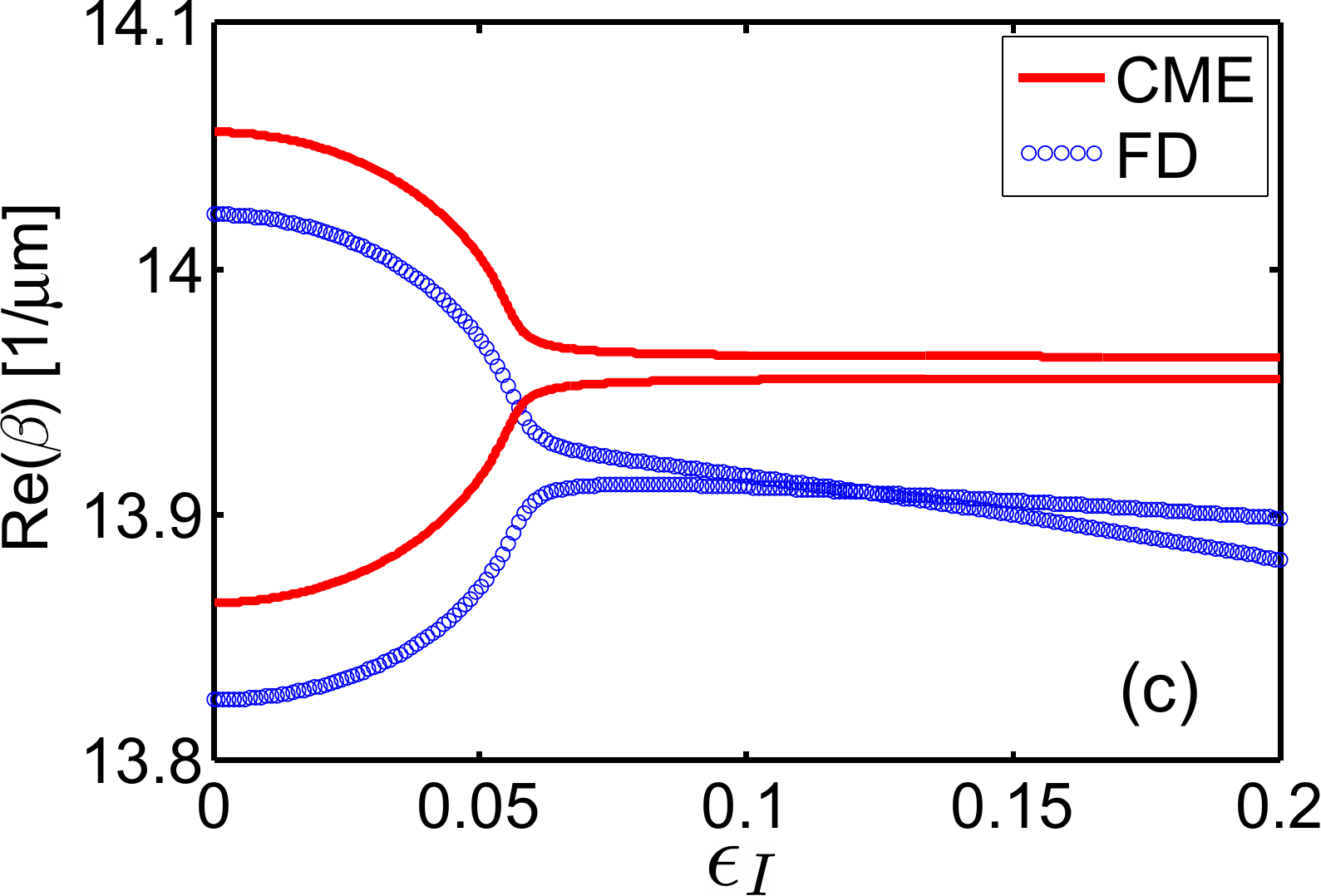}
	\includegraphics[width = 0.49\columnwidth, clip=true, trim = {0	0 0 0}]{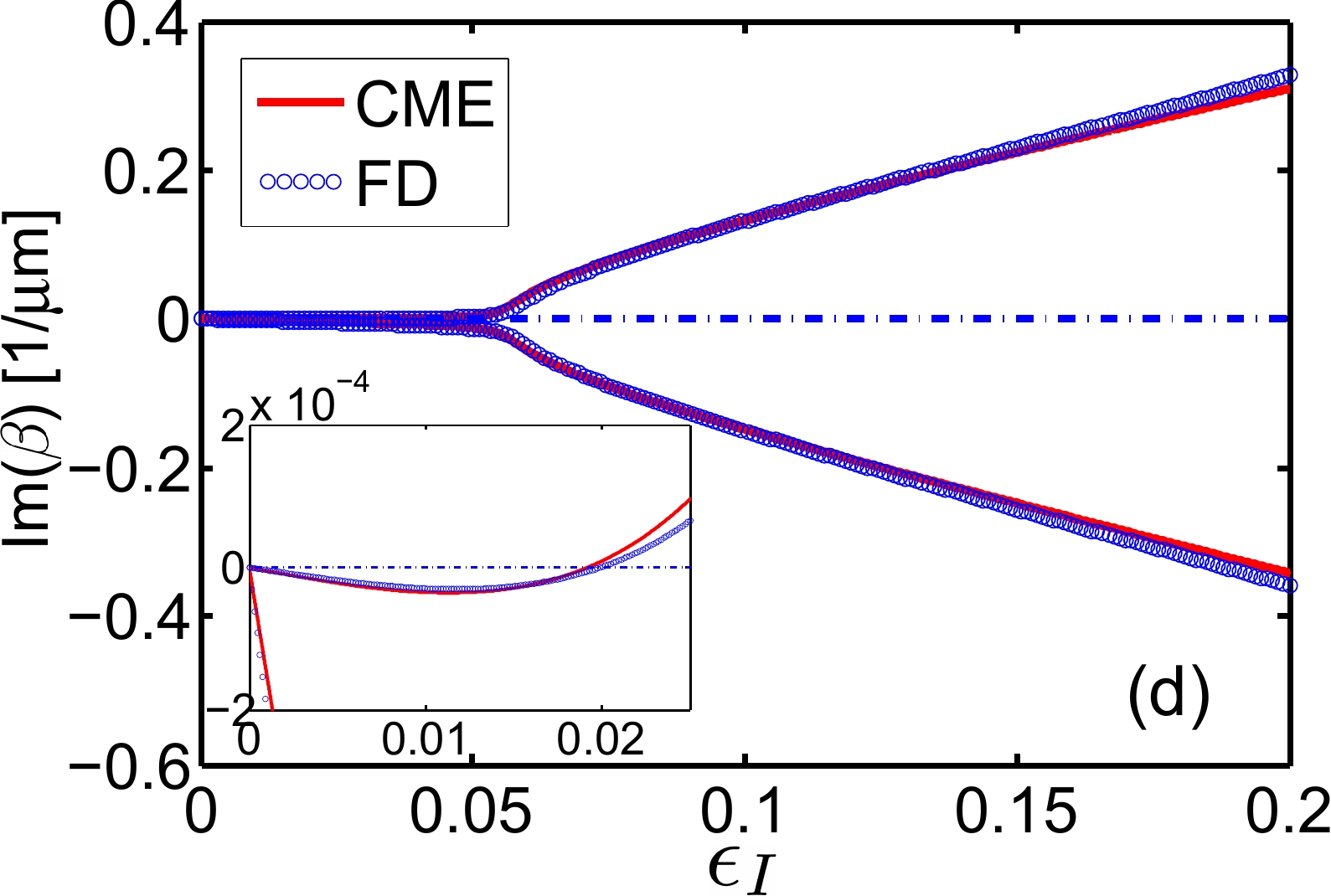}\\
	\caption{Dependence of the real (a), (c) and imaginary (b), (d) parts of the propagation constants of coupler modes on the gain/loss amplitude $\epsilon_I$ for (a), (b) m-$\mathcal{PT}$ coupler with $\beta_1 \neq \beta_2$ and (c), (d) modified m-$\mathcal{PT}$ coupler where $\beta_1 = \beta_2$. In subplots (b), (d), the zero level is denoted by a blue dashed line and the insets show zooms on the region of low $\epsilon_I$.}
	\label{fig:curves}
\end{figure}

In the m-$\mathcal{PT}$ systems, the exceptional point cannot be clearly identified. Nevertheless, we show here that at least one of the modes of the system shows the behavior similar to the transition from full to broken $\mathcal{PT}$ symmetry. Figures~\ref{fig:curves}(a) and (b) show the evolution of the propagation constants of the modes of our coupler as a function of gain/loss amplitude $\epsilon_I$. The imaginary part of permittivity is given by $\epsilon_{1,I} = -\epsilon_I$, $\epsilon_{2,I} = M\epsilon_I$, and $M$ chosen in such a way that $\Im m(\beta_+) = 0$ at $\epsilon_I = 0.02$.

\Cref{fig:curves}(b) shows that the imaginary part of the propagation constant of the fundamental mode $\Im m(\beta_+)$ remains small for $\epsilon_I<0.05$. Above this value, $\Im m(\beta_+)$ starts to grow rapidly indicating that the mode experiences gain. As seen from the inset in \cref{fig:curves}(b) at $\epsilon_I=0.02$ the propagation constant  $\Im m(\beta_+) = 0$, as intended by the design. For $\epsilon_I<0.02$, the fundamental mode experiences loss with the decay rate~\cite{zolla2005foundations} $\alpha = 20 \Im m(\beta_+)/\ln(10) \le 0.4$~dB/mm, which is negligible in integrated systems where the typical propagation length is of the order of hundreds of micrometers. The qualitative behavior of the propagation constant of the fundamental mode of an m-$\mathcal{PT}$ coupler is similar to the one observed in $\mathcal{PT}$-symmetric systems. On the contrary, the second-order mode, for which the system was not optimized, experiences loss for all values of $\epsilon_I$.

Similar to the $\mathcal{PT}$-symmetric system, the real parts of the propagation constants of a m-$\mathcal{PT}$ system become closer to each other with the increase of $\epsilon_I$. Although here they do not merge into one curve, due to the asymmetry of the waveguides. 

The m-$\mathcal{PT}$ system discussed above was designed so that the modes of isolated waveguides building the coupler have different propagation constants $\beta_1 \neq \beta_2$. Here, we show that for asymmetric m-$\mathcal{PT}$ couplers ($\Re e[\epsilon(x)] \neq \Re e[\epsilon(-x)] $) built of waveguides where $\beta_1 = \beta_2$ the diagrams $\beta_{\pm}(\epsilon_I)$ resemble the standard diagrams for $\mathcal{PT}$-symmetric systems. Figures~\ref{fig:curves}(c) and (d) show the $\beta_{\pm}(\epsilon_I)$ diagrams for the system with the width of the gain waveguide ($d =0.4$~$\mu$m) chosen in such a way that $\beta_1 = \beta_2$. We observe that the real parts of the propagation constants are much closer than in the case of the original m-$\mathcal{PT}$ system discussed above, even though they still do not merge to a single value as in the case of the $\mathcal{PT}$-symmetric systems. Also the imaginary parts of the propagation constants of both coupler modes show behavior similar to a $\mathcal{PT}$-symmetric system, as they  remains close to zero for $\epsilon_I<0.05$. Nevertheless, as seen in the inset of \cref{fig:curves}(d), this system allows for strictly lossless propagation of only one (fundamental) mode. The relative difference between the real parts of propagation constants computed using CME and FD methods is smaller than~1\%. However, a qualitative difference between the results of the two methods can be observed in modified m-$\mathcal{PT}$ system with $\beta_1  = \beta_2$. The plots of real parts of the coupler mode propagation constants obtained using the FD method cross at $\epsilon_I \approx 0.12$. This interesting phenomenon will be discussed in detail elsewhere.

In conclusion, we have shown that m-$\mathcal{PT}$ couplers, whose permittivity distribution does not fulfill the $\mathcal{PT}$ condition, can support a mode with a real propagation constant. In such asymmetric non-Hermitian couplers, a proper choice of the ratio between gain and loss results in a stable mode propagation and energy conservation. Moreover, the ratio between the light intensity in the gain and loss waveguide at the output of the coupler does not depend on the type of the input field. This ratio can be designed arbitrarily by changing geometric and materials parameters of the coupler, therefore offering a possibility to design a beam splitter with arbitrary split ratio that is independent of the input beam. Another feature distinguishing our m-$\mathcal{PT}$ system from conventional $\mathcal{PT}$-symmetric couplers is that the analogue of the exceptional point can be identified only for one of the coupler modes and the real eigenvalue can only be obtained at a single specifically chosen pair of gain and loss values. Below this point, the chosen mode experiences loss that is negligible at the typical propagation distances in integrated optics. For gain and loss values above this point, this mode experiences gain, similar to the $\mathcal{PT}$-symmetric system. Finally, we demonstrate that in a particular case of asymmetric coupler, designed such that the modes of each waveguide composing the coupler have equal propagation constants, the behavior of both super-modes supported by the coupler resembles that of the $\mathcal{PT}$-symmetric systems. These results are likely to have an impact for both practical realization of $\mathcal{PT}$-like behavior in laboratory experiments and more generally, in the basic science and applications of  integrated optics and topological photonics.


Authors thank Dr.~Liang Feng from University at Buffalo, The State University of New York for helpful discussions. This work was supported by US Army Research Office Award \#W911NF-15-1-0152.


%

\end{document}